\documentclass[cits]{PoS}
\usepackage{amsmath,amssymb}
\usepackage{psfig}
\usepackage{graphicx}
\title{Light Meson Distribution Amplitudes}

\ShortTitle{Light Meson Distribution Amplitudes}

\author{R.\ Arthur$^a$, P.A.\ Boyle$^a$, D.\ Br\"ommel$^b$, M.A.\ Donnellan$^d$,
  J.M.\ Flynn$^c$, A.\ J\"uttner$^e$, ~~~~~~ H.\ Pedroso de Lima$^c$, \speaker{T.D.\ Rae}$^c$, C.T.\ Sachrajda$^c$ and B.\ Samways$^c$\\
  \llap{$^a$} School of Physics and Astronomy, University of Edinburgh,
  Edinburgh EH9 3JH, UK\\
  \llap{$^b$} Institute for Advanced Simulation, J\"ulich Supercomputing Centre, Forschungszentrum J\"ulich GmbH,
  52425 J\"ulich, Germany\\  
  \llap{$^c$} School of Physics and Astronomy, University of Southampton,
  Southampton SO17 1BJ, UK\\
  \llap{$^d$} Deutsches Elektronen-Synchrotron DESY, 15738
  Zeuthen, Germany\\
  \llap{$^e$} CERN, Physics Department, 
  1211 Geneva 23, Switzerland\\
  E-mail: \email{t.d.rae@phys.soton.ac.uk}}

\abstract{We calculated the first two moments of the light-cone distribution amplitudes for the pseudoscalar mesons ($\pi$ and $K$) and the longitudinally polarised vector mesons ($\rho$, $K^*$ and $\phi$) as part of the UKQCD and RBC collaborations' $N_f=2+1$ domain-wall fermion phenomenology programme. These quantities were obtained with a good precision and, in particular, the expected effects of $SU(3)$-flavour symmetry breaking were observed. Operators were renormalised non-perturbatively and extrapolations to the physical point were made, guided by leading order chiral perturbation theory. The main results presented are for two volumes, $16^3\times 32$ and $24^3\times 64$, with a common lattice spacing. Preliminary results for a lattice with a finer lattice spacing, $32^3\times64$, are discussed and a first look is taken at the use of twisted boundary conditions to extract distribution amplitudes.}

\FullConference{The XXVIII International Symposium on Lattice Field Theory, Lattice2010\\
		June 14-19, 2010\\
		Villasimius, Italy}

\begin{document}

\section{Introduction}
These proceedings will give our current values for the lowest moments of the leading twist meson distribution amplitudes (DAs) \cite{{Boyle:2006pw},{Donnellan:2007xr},{Boyle:2008nj}} on the previously analysed $24^3$ and $16^3$ ensembles, for which the results can be considered finalised. These will be presented in a forthcoming RBC/ UKQCD paper \cite{Arthur:2010}. Preliminary results for the finer $32^3$ ensemble will then be presented, with a continuum extrapolation in mind. Finally a first look at the use of partially twisted boundary conditions for the extraction of distribution amplitudes will be presented.

The calculations were performed on lattices of sizes $16^3\times32\times16$ and $24^3\times64\times16$ with a common lattice spacing $a^{-1}=1.733(25)$GeV and $32^3\times64\times16$ with $a^{-1}=2.2856(28)$GeV. The gauge field ensembles are drawn from joint RBC/UKQCD datasets using $N_f=2+1$ flavours of domain wall fermions (DWF) and an IWASAKI gauge action. The last dimension, of length $16$, for each lattice is the extra, $5^{\mathrm{th}}$, dimension required to implement DWF. The light quark masses range from $0.005$ to $0.03$ (pion masses $331~$MeV to $672~$MeV) with a fixed strange quark mass, $am_s=0.04$, for the $24^3$ ensemble. For the $32^3$ ensemble, the light quark mass ranges from $0.004$ to $0.006$ (pion masses $290~$MeV to $420~$MeV) with a unitary $am_s=0.03$ and partially quenched $am_s=0.025$ \cite{{Allton:2008pn},{Aoki:2010}}.
\begin{table}[h]
\begin{center}
\begin{tabular}{l l l l l l}
\hline
\hline
Ensemble & $am_q$ & Range & $N_{\mathrm{meas}}$ & $t_{\mathrm{src}}$ locations & Smearing \\
\hline
$16^3\times32$ & 0.005 & 900-4480 & 180 & 0,32,16 & HL-HL\\
$~$ & 0.01 & 800-3940 & 315 & 0,32 & GL-GL\\
$~$ & 0.02 & 1800-3580 & 90 & 0,32 & HL-HL\\
$~$ & 0.03 & 1260-3040 & 90 & 0,32 & HL-HL\\
\hline
$24^3\times32$ & 0.01 & 500-3990 & 350 & 0,8,16,24 & GL-GL\\
$~$ & 0.02 & 500-3990 & 350 & 0,8,16,24 & GL-GL\\
$~$ & 0.03 & 4030-7600 & 358 & 0,16 & GL-GL\\
\hline
$32^3\times32$ & 0.004 & 760-2410 &  86 & 0,16,32,48 & LL-LL\\
$~$ & 0.006 & 500-3240 & 82 & 0,16,32,48 & LL-LL\\
$~$ & 0.008 & 500-2960 & 61 & 0,16,32,48 & LL-LL\\
\hline
\hline
\end{tabular}
\caption{Parameters for the 3 datasets. $N_{\mathrm{meas}}$ is the number of measurements for each source position $t_{\mathrm{src}}$. The total number of measurements is then $N_{\mathrm{meas}}\times N_{\mathrm{src}}$. In the smearing column XY-XY denotes the contraction of two quark propagators with X-type smearing at source and Y-type smearing at sink: G=Gaussian wavefunction, H=Hydrogen wavefunction, L=point.}
\end{center}
\end{table}
\vspace{-0.8cm}
\section{Meson Distribution Amplitudes}
Distribution amplitudes are introduced in the QCD description of hard exclusive processes. They encode the non-perturbative QCD effects that occur from factorisation and are important for form factors at large $q^2$ and also for B-decays. They are universal hadronic properties that do not depend on the process itself.

The leading twist DAs for pseudoscalar and vector mesons are defined via vacuum-to-meson matrix elements of quark-antiquark light-cone operators
\begin{eqnarray}
\langle 0|\bar{q}(z)\gamma_\rho\gamma_5 \mathcal{P}(z,-z)q(-z)|\pi(p)\rangle\bigl|_{z^2=0}&\equiv& f_\pi(ip_\rho)\int^1_0\textrm{d}u~e^{i(u-\bar{u})p.z}\phi_\pi(u,\mu),\\
\langle 0|\bar{q}(z)\gamma_\mu\gamma_5 \mathcal{P}(z,-z)q(-z)|\rho(p;\lambda)\rangle\bigl|_{z^2=0}&\equiv& f_\rho m_\rho p_\mu\frac{\epsilon_{(\lambda)}.z}{p.z}\int^1_0\textrm{d}u~e^{i(u-\bar{u})p.z}\phi_\rho^{||}(u,\mu).
\end{eqnarray}
The path ordered exponential $\mathcal{P}$ ensures gauge invariance and the DAs are normalised to one when integrated over the momentum fraction $u$ carried by the quarks ($\bar{u}=1-u$). It is useful to parameterise the DAs via their moments,
\begin{equation}
\langle\xi^n\rangle_\pi(\mu)=\int_{0}^1\textrm{d}u \xi^n\phi_\pi(\xi,\mu)
\end{equation}
where $\xi=u-\bar{u}$ is the difference between the momentum fractions. 

These moments appear in matrix elements of local operators with $n$ derivatives. The bare moments (denoted $\langle\xi^n\rangle^{\mathrm{bare}}$ where $n=1,2$) are accessible on the lattice through ratios of two-point correlation functions. Taking the simplest example of the first moment pseudoscalar, for large Euclidean times $t$ and $T-t$
\begin{equation}\label{eq1}
R^P_{\{\rho\mu\};\nu}(t,p)\equiv\frac{\sum_{x}e^{ip.x}\langle0|\mathcal{O}_{\{\rho\mu\}}(t,x)P^\dagger(0)|0\rangle}{\sum_{x}e^{ip.x}\langle0|A_\nu(t,x)P^\dagger(0)|0\rangle}\stackrel{\mathrm{Large~}t}{=}\frac{ip_\rho p_\mu}{p_\nu}\langle\xi^1\rangle 
\end{equation}
\begin{equation}\label{eq2}
R^P_{\{\rho\mu\nu\};\sigma}(t,p)\equiv\frac{\sum_{x}e^{ip.x}\langle0|\mathcal{O}_{\{\rho\mu\nu\}}(t,x)P^\dagger(0)|0\rangle}{\sum_{x}e^{ip.x}\langle0|A_\sigma(t,x)P^\dagger(0)|0\rangle}\stackrel{\mathrm{Large~ }t}{=}\frac{ip_\rho p_\mu p_\nu}{p_\sigma}\langle\xi^2\rangle 
\end{equation}
where $A$ and $P$ are the axial and pseudoscalar quark bilinears. The operator with $n$ derivatives is given by
\begin{equation}
\mathcal{O}_{\{\mu\mu_1...\mu_n\}}(x,t)\equiv \bar{q}(x,t)\gamma_{\{\mu}\overleftrightarrow{D}_{\mu_1}...\overleftrightarrow{D}_{\mu_n\}}q'(x,t)
\end{equation}
where the braces in the subscript indicate symmeterisation of the enclosed indices and subtraction of traces.

It is desirable to choose indices so that operator mixing is kept under control and so that there are as few non-zero momentum components as possible~\cite{Arthur:2010}. We obtain the first moment from $R^P_{\{\rho 4\};4}(t,p)$ (where the index 4 corresponds to the time direction) with $\rho=1,2$ or $3$ and a single non-zero component of momentum, $|p_\rho|=2\pi/L$. Similarly the second moment is extracted from $R^P_{\{\rho \mu 4\};4}(t,p)$ with at least two non-zero components of momentum, taking $\rho$, $\mu=1,2$ or $3$ with $\rho \neq \mu$ and $|p_\rho|=|p_\mu|=2\pi/L$. A similar method is used to extract moments for the polarised vector mesons.

\section{$\langle\xi^1\rangle$ Results}
For the first moment of the kaon DA, chiral perturbation theory predicts that the extrapolation to the chiral limit is linear in $m_s-m_q$ \cite{Chen:2003fp}. Our data clearly shows the expected SU(3) symmetry breaking effects for all three ensembles. The bare results are plotted in Fig.~1 as a function of the mass of the light quark, where $m_s a=0.04$ for the $16^3$ and $24^3$ ensembles and $m_s a=0.03$ (triangular points, Fig.~1) and partially quenched $m_s a=0.025$ (circular points, Fig.~1) for the $32^3$ ensemble. We see a similar linear behaviour for $\langle \xi^1 \rangle_{K^*}$ and so perform the same extrapolation. For the $K^*$ we see a hint of a finite volume effect, however, where there are results for both volumes at the same light-quark mass, they agree within statistical uncertainties. We extrapolate to the physical points $a(m_s-m_q)=0.0362(16)$ ($16^3$ and $24^3$) and $0.02816(80)$ ($32^3$) \cite{{Allton:2008pn},{Aoki:2010}} for both $\langle \xi^1 \rangle_K$ and $\langle \xi^1 \rangle_{K^*}$. The results are given in Table~2.

\begin{figure}
\centering
\begin{tabular}{cc}
\psfig{file=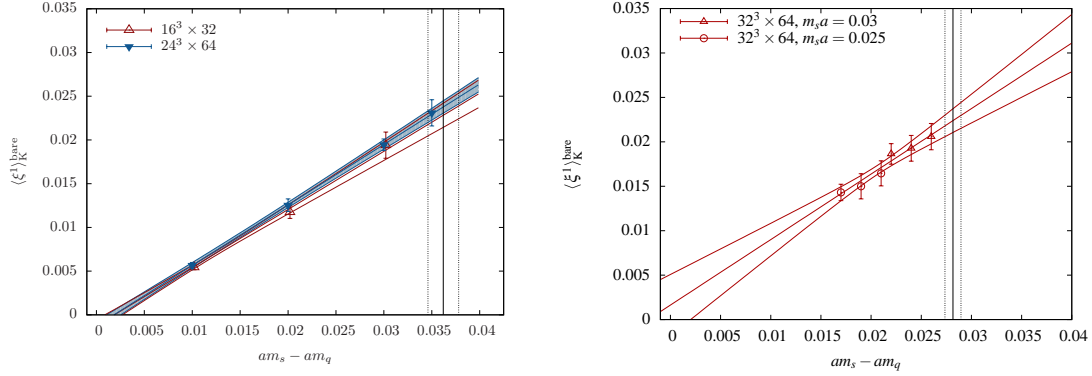,width=0.48\linewidth} &
\scalebox{0.57}{\input{K1st0608.tex}} \\
\end{tabular}
\caption{Linear extrapolation for the first moment $K$. The solid vertical lines mark the physical point and the dashed lines show the error on this point. The left hand plot is $16^3$ and $24^3$ and the right hand plot is $32^3$ where the circular points are partially quenched.}
\end{figure}

\section{$\langle\xi^2\rangle$ Results}
For the second moments we have some guidance from chiral perturbation theory; that there should be no non-analytic dependence at 1 loop and we should fit linearly in $m_\pi^2$ \cite{Chen:2005js}. We see no obvious dependence on the quark mass and so perform a linear extrapolation in $m_l$ at a fixed simulated $m_s$. The bare results are plotted in Fig.~2. We evaluate the bare second moment values at the chiral limit and present these in Table~2. For the second moments we see no obvious indication of finite size effects.
\begin{table}[h]
\begin{center}
\begin{tabular}{@{}l l l l l l l l@{}}
\hline
\hline
$~$ & $\langle\xi^2\rangle_\pi^{\mathrm{bare}}$ & $\langle\xi^1\rangle_K^{\mathrm{bare}}$ & $\langle\xi^2\rangle_K^{\mathrm{bare}}$ & $\langle\xi^1\rangle_{K^*}^{||\mathrm{bare}}$ & $\langle\xi^2\rangle_{K^*}^{||\mathrm{bare}}$ & $\langle\xi^2\rangle_\rho^{||\mathrm{bare}}$ & $\langle\xi^2\rangle_\phi^{||\mathrm{bare}}$ \\
\hline
$\small{16^3}$ & \small{0.112(5)} & \small{0.0228(14)(11)}   & \small{0.112(4)} & \small{0.02443(96)(107)} & \small{0.110(6)} & \small{0.109(10)} & \small{0.107(5)}\\
$\small{24^3}$ & \small{0.125(7)} & \small{0.02377(71)(100)} & \small{0.117(5)} & \small{0.0281(13)(14)}   & \small{0.118(7)} & \small{0.118(7)}  & \small{0.107(4)}\\
\hline
$\small{32^3}$ & \small{0.17(2)}& \small{0.022(1)(1)}& \small{0.14(1)}&\small{0.034(2)(1)}& \small{0.12(2)}&\small{0.15(3)} &\small{0.12(1)}\\
\hline
\hline
\end{tabular}
\caption{Results for the bare values of moments of the distribution amplitudes. The errors are statistical and (for the first moment) due to the uncertainty in the physical point for the chiral extrapolation. Note that bare results at different lattice spacings should not be compared directly (as the renormalisation constants depend on the lattice spacing).}
\end{center}
\end{table}
\begin{figure}
\centering
\begin{tabular}{cc}
\psfig{file=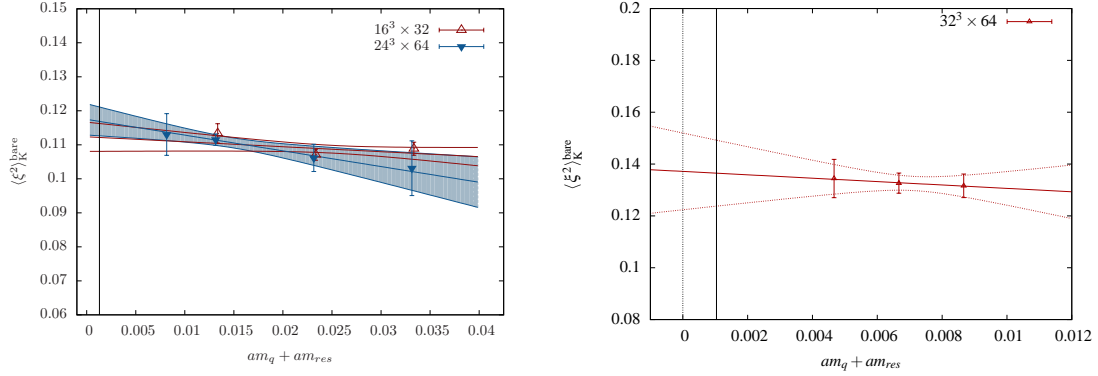,width=0.48\linewidth} &
\scalebox{0.57}{\input{K2nd0608.tex}} \\
\end{tabular}
\caption{Linear extrapolation for the second moment for the $K$ meson. The solid vertical lines mark the physical point. The Left hand plot is $16^3$ and $24^3$ and the right hand plot is $32^3$ using the unitary strange quark mass.}
\end{figure}
\section{Renormalisation of results}
We use the Rome-Southampton RI$^\prime$/MOM scheme to renormalise the relevant operators non-perturbatively \cite{{Boyle:2008nj},{Arthur:2010},{Martinelli:1994ty}}. Multiplying the bare moments by the renormalisation factors, we obtain the renormalised results in $\overline{\textrm{MS}}$ at $\mu=2$~GeV. It should be noted that the $16^3$ and $24^3$ lattices have different renormalisation factors to the $32^3$.

The $16^3$ and $24^3$ results agree with the $32^3$ results within or just beyond $1\sigma$ (apart from the $K^*$ first moment). We estimate a formal discretisation error, of $O(a^2\Lambda^2_{QCD})$ corresponding to $\sim 4\%$ ($\sim 2\%$) for the $16^3$ and $24^3$ ($32^3$) ensembles from the $O(a)-$improved DWF action and operators, which is included in Table~3. The $32^3$ results are preliminary and further work will be done to confirm its values and errors. Following this, a continuum extrapolation will be made to better estimate any discretisation effects.
\begin{table}[h]
\begin{center}
\begin{tabular}{l l l l l l l l}
\hline
\hline
$~$ & $\langle\xi^2\rangle_\pi$ & $\langle\xi^1\rangle_K$ & $\langle\xi^2\rangle_K$ & $\langle\xi^1\rangle_{K^*}^{||}$ & $\langle\xi^2\rangle_{K^*}^{||}$ & $\langle\xi^2\rangle_\rho^{||}$ & $\langle\xi^2\rangle_\phi^{||}$ \\
\hline
$\small{16^3}$ & \small{0.25(1)(2)} & \small{0.035(2)(2)}   & \small{0.25(1)(2)} & \small{0.037(1)(2)} & \small{0.25(1)(2)} & \small{0.25(2)(1)} & \small{0.24(1)(1)}\\
$\small{24^3}$ & \small{0.28(1)(2)} & \small{0.036(1)(2)} & \small{0.26(1)(2)} & \small{0.043(2)(3)}   & \small{0.25(2)(1)} & \small{0.27(1)(2)}  & \small{0.25(2)(1)}\\
$\small{32^3}$ & \small{0.36(5)(2)}& \small{0.034(2)(2)}& \small{0.30(3)(2)}&\small{0.052(3)(3)}& \small{0.26(4)(2)}&\small{0.32(8)(2)} &\small{0.26(2)(2)}\\
\hline
\hline
\end{tabular}
\caption{Results for the renormalised values of the distribution amplitudes in $\overline{\textrm{MS}}$ at $\mu=2$~GeV. The first error is statistical and the second includes systematic errors from $m_s$, discretisation and renormalisation.}
\end{center}
\end{table}
\section{Partially twisted boundary conditions}
We used the Z4PSs4 and Z4PSs3 datasets from the RBC/UKQCD $K \rightarrow \pi$ form factor runs, where the naming convention is as follows: spin-diluted $\mathbb{Z}(2)\times \mathbb{Z}(2)$ noise source and point sink with strange quark mass $am_s=0.04$ and $am_s=0.03$ respectively. Both sets have $am_q=0.005$ and consist of 1180 measurements \cite{Boyle:2010bh}. The gauge field configurations are generated through combining sea quarks obeying periodic boundary conditions with valence quarks that have twisted boundary conditions \cite{Sachrajda:2004mi}. The valence quarks therefore satisfy
\begin{equation}
\psi(x_k+L)=e^{i\theta_k}\psi(x_k),~~~k=1,2,3,
\end{equation}
where $\psi$ is either a strange quark or one of the light quarks. 
 
We calculate the correlation functions on these datasets with zero fourier momentum and we look at cases where only one of the valence quarks is twisted. Therefore the kaon's momentum is induced purely by the twist angle of the valence quark. The twist angle is only along one of the spatial directions, which is changed regularly in order to reduce correlations. The twist angles used for this analysis are $\theta_l=1.600$ for the light quark for both strange quark masses and $\theta_s=2.5087(2.7944)$ for $am_s=0.03(0.04)$ \cite{Boyle:2010bh}. The momentum of the meson is then, ${\bf p}={\bf \theta}/ L$.
 
\begin{figure}[h]
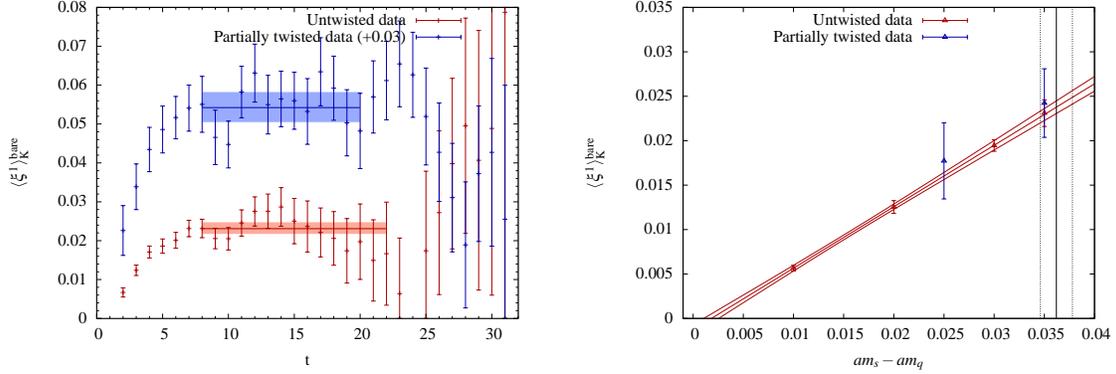

\centering
\begin{tabular}{cc}
\scalebox{0.57}{\input{K1stratio24twist.tex}} &
\scalebox{0.57}{\input{K1st24twist.tex}}\\
\end{tabular}
\caption{Comparison of using twisted boundary conditions to induce momentum with just using the Fourier momentum alone, for the kaon first moment, $24^3$. The partially twisted data shown on the left hand plot is for the $0.04$ mass strange quark, for which the data-points have been shifted by ($+0.03$) in the y-direction. The partially twisted data on the right hand plot shows the results for both the $0.04$ and $0.03$ mass strange quark.}
\end{figure}

Fig.~3 compares the partially twisted data with the untwisted data presented earlier in this paper. We see that the results agree well and that using twisted boundary conditions to extract the moments of PDAs (at least for the kaon first moment) is possible. It should be noted that we only show the cases where the strange quark is twisted. The signal where we twist the light quark is too weak to allow a fit. The ratio from which we extract the first moment, see Eq~\ref{eq1}, is proportional to the momentum and hence the twist angle. The twist is small for the light quark and leads to a poor signal for the first moment.

\section{Summary}

We have computed the lowest two non-vanishing moments of the distribution amplitudes for the $\pi$, $K$, $K^*$, $\rho$ and $\phi$ mesons, using non-perturbative renormalisation, for two ensembles with a common lattice spacing but with different volumes and also for a third ensemble with a finer lattice spacing. The renormalised results are presented in Table~3. We do not see any finite size effects within errors. The $32^3$ results are preliminary and a further investigation will be done to confirm its values and errors. We have shown promising results for the extraction of the kaon's first moment using partially twisted boundary conditions.

\section{Acknowledgements}
The calculations reported here used the QCDOC computers at Edinburgh University, Columbia University and Brookhaven National Laboratory (BNL). We thank the University of Southampton for access to the Iridis computer system used in the calculation of the non-perturbative renormalisation factors and also in extending the QCDOC pda runs for the $0.008$ and $0.006$ strange quark masses on the $32^3$ ensemble. DB, MAD, JMF, AJ, TDR, CTCS and BS acknowledge support from STFC Grant ST/G000557/1 and from EU contract MRTN-CT-2006-035482 (Flavianet); RA and PAB from STFC grant ST/G000522/1; PAB from an RCUK Fellowship.


\end{document}